\documentclass[twocolumn,showpacs,aps,prl,letterpaper]{revtex4}
\usepackage{graphicx}
\usepackage{dcolumn}
\usepackage{amsmath}
\usepackage{epsfig}
\usepackage{hyperref}
\long\def\inst#1{\par\nobreak\kern 4pt\nobreak
    {\itshape #1}\par\vskip 10pt plus 3pt minus 3pt}
\RequirePackage{xspace}
\usepackage{relsize}

\begin{document}

\title{\large \bfseries \boldmath Spectrum of Higher excitations of $B$ and $D$ mesons in the relativistic potential model}
\author{Jing-Bin Liu}\email{liujingbin077@mail.nankai.edu.cn}
\author{Mao-Zhi Yang}\email{yangmz@nankai.edu.cn}
\affiliation{School of Physics, Nankai University, Tianjin 300071,
P.R. China}

\date{\today}

\begin{abstract}
We study the masses and wave functions of heavy-light quark-antiquark system in the relativistic potential model,
where we focus on the higher excited states of $B$ and $D$ mesons.  The resonances of $D_J(2740)$, $D_J^*(2760)$, $D_J^*(3000)$ and $D_{sJ}^*(2860)$ recently observed in experiments are considered. We find that the four resonances can be identified as the $|1^{1,3}D_2\rangle$, $|1^3D_3\rangle$, $|1^{1,3}F_3\rangle$ and $|1^3D_3\rangle$ states due to the prediction to the states with $l=2$ and $l=3$ in the potential-model calculation. The masses of $n=2$ excited  states are also calculated. All the observed mesons can be accommodated successfully in the potential model.
\end{abstract}


\maketitle

\section*{I Introduction}  

Heavy-light quark-antiquark system $Q\bar{q}$ plays an important role in studying the strong interactions between quark and antiquarks. Great progress was made in the measurement of the spectrum of heavy-light quark-antiquark system in the last decade. In the charm sector, several new excited meson states were discovered in the previous several years \cite{Ds2632,Ds2860,Ds2700,Ds2009,D2010,LHCb}. For $D_J$ mesons, the resonances of $D(2740)^0$, $D^*(2760)^{0,+}$ \cite{D2010}, $D^*(2650)$ and $D^*(3000)$ \cite{LHCb} were observed, while for $D_{sJ}$ mesons, the resonances $D_{sJ}^+(2632)$ \cite{Ds2632}, $X(2690)$, $D_{sJ}^+(2860)$ \cite{Ds2860},  $D_{sJ}^+(2700)$ \cite{Ds2700} and $D_{sJ}^+(3040)$ \cite{Ds2009} were established. The spectroscopy provides a powerful test of the theoretical predictions based on the quark model in the standard model.

The bound state of the heavy-light quark-antiquark system can be studied with the relativistic wave equation, where an effective potential compatible with QCD is taken \cite{RGG,GI}. The potential shows a linear confining behavior at long distance and a Coulombic behavior at short distance. The properties of $B$ and $D$ mesons have been investigated in the relativistic potential
model in Refs. \cite{GI,RP1,RP2,RP3,RP4,RP5,ymz}. In our recent work \cite{LY}, the spectra for lower orbital and radial excitations of $B$ and $D$ mesons were calculated, a good agreement with experiment \cite{PDG} was obtained. Here we extend our previous work to include the higher excited states and compare the prediction with the spectrum measured in experiment in recent years. The excited states of the heavy-light system have also been studied in Ref. \cite{PE} in the relativistic quark model, which includes the leading order corrections in $1/m_{c,b}$ expansion. The Hamiltonian used in Ref. \cite{PE} is a kind of relativistic Dirac Hamiltonian. Compared with Ref. \cite{PE}, the method used in this work is different and the recently observed states are also carefully studied in this work.

The organization of this paper is as follows. In section II, we give the effective Hamiltonian for the heavy-light quark-antiquark system and solve the relativistic wave equation theoretically. Section III is for the numerical result and discussion. In section IV we have a brief summary.

\section*{II The effective Hamiltonian and the solution of the wave equation }
In the potential model one assumes that hadrons can be treated in terms of valence-quark configurations.  Here we employ the relativistic potential model to describe the heavy-light $B$ and $D$ mesons. The effective potential between the quark and antiquark in the meson is one-gluon-exchange (OGE) dominant at short distances and with a linear confinement at long distances. Since there is a light quark in the heavy-light system, it requires to include relativistic dynamics in the Hamiltonian. The relativistic version of Schr\"{o}dinger wave equation describing the heavy-light system is written as
\begin{equation}
H\psi(\vec{r})=E\psi(\vec{r}). \label{e1}
\end{equation}
In the rest frame of the bound state system, the eigenvalues of the Schr\"{o}dinger equation will be the masses of mesons.

The effective Hamiltonian can be written as
\begin{equation}
H=H_0+H^\prime, \label{e2}
\end{equation}
with
\begin{equation}
H_0=\sqrt{\vec{p}_1^2+m_1^2}+\sqrt{\vec{p}_2^2+m_2^2}+V(r),
\end{equation}
where $m_1$, $\vec{p}_1$ are the mass and momentum of the heavy quark, respectively, and
$m_2$, $\vec{p}_2$ the mass and momentum of the light antiquark. $V(r)$ is the funnel potential, which includes only the spin-independent part of the strong interaction between the heavy and light quarks
\begin{equation}
V(r)=-\frac{4}{3}\frac{\alpha_s(r)}{r}+b\;r+c.\label{e4}
\end{equation}
The potential is taken as a combination of a Coulomb term and a linear confining term, whose behavior is compatible with QCD at both short- and long-distance \cite{GI,Cornell1,Cornell2}. The first term is obtained from the calculation of one-gluon-exchange diagram  in perturbative QCD. It dominates at short-distance. The second term is the linear confining
term. The third term is a phenomenological constant, which can be adjusted to give the correct ground
state energy-level of the quark-antiquark system.

The running coupling constant $\alpha_s(r)$ in eq.(\ref{e4}) is obtained from the coupling
constant in momentum space $\alpha_s(Q^2)$ after the Fourier transformation. It can be written in a more convenient form \cite{GI}
\begin{equation}
\alpha_s(r)=\sum_i\alpha_i\frac{2}{\sqrt{\pi}}\int_0^{\gamma_i
r}e^{-x^2}dx,
\end{equation}
where $\alpha_i$ and $\gamma_i$ are free parameters which can be fitted to make the behavior of the running coupling constant at
short distance be consistent with the coupling constant in momentum space predicted by QCD. In this work, we take $\alpha_1=0.15$, $\alpha_2=0.15$, $\alpha_3=0.20$, and $\gamma_1=1/2$, $\gamma_2=\sqrt{10}/2$, $\gamma_3=\sqrt{1000}/2$ \cite{ymz}.

The second term $H^\prime$ in eq.(\ref{e2}) is the spin-dependent part of the Hamiltonian
\begin{equation}
H^\prime=H^{\rm hyp}+H^{\rm so},
\end{equation}
where $H^{\rm hyp}$ is the spin-spin hyperfine-interaction term, which is
\begin{eqnarray}
H^{\rm hyp}=&&\frac{32\pi}{9m_1 \tilde{m}_{2a}}\alpha_s(r)\delta_\sigma(r)\vec{s}_1\cdot\vec{s}_2
\nonumber\\
&&+\frac{4}{3}\frac{\alpha_s(r)}{m_1\tilde{m}_{2b}}\frac{1}{r^3}\left(\frac{3\vec{s}_1\cdot \vec{r}\vec{s}_2\cdot \vec{r}}{r^2}-
\vec{s}_1\cdot\vec{s}_2\right) \label{e7}
\end{eqnarray}
with
\begin{equation}
\delta_\sigma (r)=(\frac{\sigma}{\sqrt{\pi}})^3e^{-\sigma^2r^2},
\end{equation}
where the parameter $\sigma$ is taken as quark mass-dependent \cite{GI}
\begin{equation}
\sigma=\sqrt{\sigma_0^2\left(\frac{1}{2}+\frac{1}{2}\left(\frac{4m_1m_2}{(m_1+m_2)^2}\right)^4\right)
+s^2_0\left(\frac{2m_1m_2}{m_1+m_2}\right)^2},
\end{equation}
here $\sigma_0$ and $s_0$ are phenomenological parameters.

The spin-orbit interaction term $H^{\rm so}$ consists of the color-magnetic term and the Thomas-precession term
\begin{eqnarray}
H^{\rm so}&=&\frac{4}{3}\frac{\alpha_s(r)}{r^3}\left( \frac{1}{m_1}+\frac{1}{\tilde{m}_{2c}}\right)
    \left( \frac{\vec{s}_1\cdot\vec{L}}{m_1}+\frac{\vec{s}_2\cdot\vec{L}}{\tilde{m}_{2c}}\right)\nonumber\\
    &-&\frac{1}{2r}\frac{\partial V(r)}{\partial r} \left( \frac{\vec{s}_1\cdot\vec{L}}{m_1^2}
    +\frac{\vec{s}_2\cdot\vec{L}}{(\tilde{m}_{2d})^2}\right),\label{e10}
\end{eqnarray}
where $\vec{L}=\vec{r}\times\vec{P}$ is the relative orbital angular momentum between the quark and antiquark.

The spin-dependent Hamiltonian $H^\prime$ is the result of OGE diagram in the non-relativistic approximation \cite{RGG,GI}. It is reasonable that there might be contributions of non-perturbative dynamics in the bound state system and relativistic corrections for the light quark. In this work, the $\delta(r)$ in the spin-spin contact hyperfine interaction term is replaced by $(\frac{\sigma}{\sqrt{\pi}})^3e^{-\sigma^2r^2}$ as in Ref. \cite{higher-charmonium}. The mass of the light quark $m_2$'s in the denominators of eqs. (\ref{e7}) and (\ref{e10}) are replaced by a set of new parameters $\tilde{m}_{2i}$, $i=a,\; b,\; c,\; d$, which is assumned to include the relativistic corrections and the bound-state effect in the heavy meson \cite{LY}. In the original expression of the spin-dependent interaction, the momenta of the quarks are dropped. However, it is not reasonable to drop them completely for the heavy-light system as the light quark should be highly relativistic. There may also be bound-state effect in the spin-dependent interaction terms which can not be described by the OGE potential. We assume that these effects can be included by replacing the light quark masses with a set of new parameters, which can be determined by fitting the spectra of heavy mesons. One can find in the next section that this assumption does work, all the masses measured in experiment can be accommodated well.

In our previous work \cite{LY}, only the states of $B$, $D$, $B_s$, $D_s$ mesons with lower orbital quantum numbers $l=0,1$ are considered, the masses and wave functions of the corresponding states with $J^P=0^-$, $1^-$, $0^+$, $1^+$, $2^+$ are obtained. In this work we extend our previous work by studying more excited states with higher orbital quantum numbers. In the heavy quark limit the spin of the heavy quark decouples from the interaction \cite{HQa1,HQa2,HQa3,HQa4,HQa5,HQa6,HQa7,HQb,GK}, therefore the spin-dependent interaction can be treated perturbatively. We choose the eigenstates of spin-independent Hamiltonian $H_0$ as the basis of perturbative expansion. The eigenstates of $H_0$ is denoted as $| JM,sl\rangle$, where $J$ is the quantum number of the total angular momentum, $M$ the magnetic quantum number, $s$ the total spin, $l$ the relevant orbital angular momentum. The $\delta_\sigma(r)$ term of the hyperfine interaction $H^{\rm hyp}$ conserves all the four quantum numbers, it gives the splitting of states with different total spin. The tensor part of $H^{\rm hyp}$ does not conserve the orbital angular momentum, it causes mixing between the states with different orbital angular momenta $^3L_J\leftrightarrow ^3L^{\prime}_J$, while the spin-orbit interaction $H^{\rm so}$ does not conserve the total spin, it causes mixing between the states with different total spin quantum numbers $^1L_J\leftrightarrow ^3L_J$.

The details of solving the wave equation have been given in \cite{ymz,LY}, which is omitted in the present paper for brevity. The readers can refer to these references.

The mass matrix is calculated perturbatively in the basis of $| JM,sl\rangle$, which can be written as
\begin{equation}
H=\left( \begin{array}{cc} H_{11}& H_{12}\\ H_{21} & H_{22} \end{array}\right ), \label{mix27}
\end{equation}
where $H_{ij}=\langle \psi_i |H|\psi_j\rangle$, with $i, j=1, 2$, $\psi_{i,j}$'s denote the eigenstates of the Hamiltonian $H_0$.
By diagonalizing the mass matrix, we can get the masses and wave functions of the mixing eigenstates. The perturbative contribution of the spin-dependent Hamiltonian $H^\prime$ to the eigenvalues of states with high orbital quantum numbers are given below (Contributions to states with lower orbital momentum have been given in our earlier work, Ref. \cite{LY}).

(1) The mass matrix of  state $J^P=2^-$

Both states with $s=0$, $l=2$ and $s=1$, $l=2$ can construct
the $J^P=2^-$ state. The basis for the mixing is $|\psi_1\rangle =|^1D_2\rangle$ and $|\psi_2\rangle =|^3D_2\rangle$.
The matrix elements of the mass matrix are
\begin{eqnarray}
H_{11}&=&E_{l=2}^{(0)}-\frac{3}{4}\langle\psi_{l=2}^{(0)}(r)| f(r)| \psi_{l=2}^{(0)}(r)\rangle ,\label{H41}\\
H_{12}&=&\sqrt{\frac{3}{2}}\langle\psi_{l=2}^{(0)}(r)|h_1(r)- h_2(r)| \psi_{l=2}^{(0)}(r)\rangle ,\label{H42}\\
H_{21}&=& H_{12}^*,\label{H43}\\
H_{22}&=& E_{l=2}^{(0)}+\langle\psi_{l=2}^{(0)}(r)| \frac{1}{4}f(r)+\frac{1}{2}g(r)\nonumber\\
&& -\frac{1}{2}h_1(r)-\frac{1}{2}h_2(r)\psi_{l=2}^{(0)}(r))\rangle .\label{H44}
\end{eqnarray}
With eqs.(\ref{H41}) $\sim$ (\ref{H44}), the cases with more $|^1D_2\rangle$ and $|^3D_2\rangle$ mixing states
can be obtained.

(2) The mass matrix of state $J^P=3^-$

The $J^P=3^-$ state is mixture of $^3D_3$ and $^3G_3$ states, both states with $s=1$, $l=2$ and $s=1$, $l=4$ can
construct the $J^P=3^-$ state. The basis for the mixing is $|\psi_1\rangle =|^3D_3\rangle$ and $|\psi_2\rangle =|^3G_3\rangle$.
The matrix elements of the mass matrix are
\begin{eqnarray}
H_{11}&=&E_{l=2}^{(0)}+\langle\psi_{l=2}^{(0)}(r)|[ \frac{1}{4}f(r)-\frac{1}{7}g(r)\nonumber\\
         &&+h_1(r)+h_2(r)]| \psi_{l=2}^{(0)}(r)\rangle ,\label{H49}\\
H_{12}&=&\frac{3}{7}\sqrt{3}\langle\psi_{l=2}^{(0)}(r)|g(r)| \psi_{l=4}^{(0)}(r)\rangle,\label{H50}\\
H_{21}&=& H_{12}^*,\label{H51}\\
H_{22}&=& E_{l=4}^{(0)}+\langle\psi_{l=4}^{(0)}(r)| [\frac{1}{4}f(r) -\frac{5}{14}g(r)\nonumber\\
    &&-\frac{5}{2}h_1(r)-\frac{5}{2}h_2(r)]| \psi_{l=4}^{(0)}(r)\rangle .\label{H52}
\end{eqnarray}
With eqs.(\ref{H49}) $\sim$ (\ref{H52}), the cases with more $|^3D_3\rangle$ and $|^3G_3\rangle$ mixing states
can be obtained.

(3) The mass matrix of  state $J^P=3^+$

Both states with $s=0$, $l=3$ and $s=1$, $l=3$ can construct
the $J^P=3^+$ state. The basis for the mixing is $|\psi_1\rangle =|^1F_3\rangle$ and $|\psi_2\rangle =|^3F_3\rangle$.
The matrix elements of the mass matrix are
\begin{eqnarray}
H_{11}&=&E_{l=3}^{(0)}-\frac{3}{4}\langle\psi_{l=3}^{(0)}(r)| f(r)| \psi_{l=3}^{(0)}(r)\rangle ,\label{H45}\\
H_{12}&=&\sqrt{3}\langle\psi_{l=3}^{(0)}(r)|h_1(r)- h_2(r)| \psi_{l=3}^{(0)}(r)\rangle ,\label{H46}\\
H_{21}&=& H_{12}^*,\label{H47}\\
H_{22}&=& E_{l=3}^{(0)}+\langle\psi_{l=3}^{(0)}(r)| \frac{1}{4}f(r)+\frac{1}{2}g(r)\nonumber\\
&& -\frac{1}{2}h_1(r)-\frac{1}{2}h_2(r)\psi_{l=3}^{(0)}(r))\rangle .\label{H48}
\end{eqnarray}
With eqs.(\ref{H45}) $\sim$ (\ref{H48}), the cases with more $|^1D_2\rangle$ and $|^3D_2\rangle$ mixing states
can be obtained.

(4) The mass matrix of state $J^P=4^+$

The $J^P=4^+$ state is mixture of $^3F_4$ and $^3H_4$ states, both states with $s=1$, $l=3$ and $s=1$, $l=5$ can
construct the $J^P=4^+$ state. The basis for the mixing is $|\psi_1\rangle =|^3F_4\rangle$ and $|\psi_2\rangle =|^3H_4\rangle$.
The matrix elements of the mass matrix are
\begin{eqnarray}
H_{11}&=&E_{l=3}^{(0)}+\langle\psi_{l=3}^{(0)}(r)|[ \frac{1}{4}f(r)-\frac{1}{6}g(r)\nonumber\\
         &&+\frac{3}{2}h_1(r)+\frac{3}{2}h_2(r)]| \psi_{l=3}^{(0)}(r)\rangle ,\label{H53}\\
H_{12}&=&\frac{\sqrt{5}}{3}\langle\psi_{l=3}^{(0)}(r)|g(r)| \psi_{l=5}^{(0)}(r)\rangle,\label{H54}\\
H_{21}&=& H_{12}^*,\label{H55}\\
H_{22}&=& E_{l=5}^{(0)}+\langle\psi_{l=5}^{(0)}(r)| [\frac{1}{4}f(r) -\frac{1}{3}g(r)\nonumber\\
    &&-3 h_1(r)-3 h_2(r)]| \psi_{l=5}^{(0)}(r)\rangle .\label{H56}
\end{eqnarray}
With eqs.(\ref{H53}) $\sim$ (\ref{H56}), the cases with more $|^3F_4\rangle$ and $|^3H_4\rangle$ mixing states
can be obtained.

(5) The mass matrix of  state $J^P=4^-$

Both states with $s=0$, $l=4$ and $s=1$, $l=4$ can construct
the $J^P=4^-$ state. The basis for the mixing is $|\psi_1\rangle =|^1G_4\rangle$ and $|\psi_2\rangle =|^3G_4\rangle$.
The matrix elements of the mass matrix are
\begin{eqnarray}
H_{11}&=&E_{l=4}^{(0)}-\frac{3}{4}\langle\psi_{l=4}^{(0)}(r)| f(r)| \psi_{l=4}^{(0)}(r)\rangle ,\label{H57}\\
H_{12}&=&\sqrt{5}\langle\psi_{l=4}^{(0)}(r)|h_1(r)- h_2(r)| \psi_{l=4}^{(0)}(r)\rangle ,\label{H58}\\
H_{21}&=& H_{12}^*,\label{H59}\\
H_{22}&=& E_{l=4}^{(0)}+\langle\psi_{l=4}^{(0)}(r)| \frac{1}{4}f(r)+\frac{1}{2}g(r)\nonumber\\
&& -\frac{1}{2}h_1(r)-\frac{1}{2}h_2(r)\psi_{l=4}^{(0)}(r))\rangle .\label{H60}
\end{eqnarray}
With eqs.(\ref{H57}) $\sim$ (\ref{H60}), the cases with more $|^1G_4\rangle$ and $|^3G_4\rangle$ mixing states
can be obtained.

(6) The mass matrix of  state $J^P=5^+$

Both states with $s=0$, $l=5$ and $s=1$, $l=5$ can construct
the $J^P=5^+$ state. The basis for the mixing is $|\psi_1\rangle =|^1H_5\rangle$ and $|\psi_2\rangle =|^3H_5\rangle$.
The matrix elements of the mass matrix are
\begin{eqnarray}
H_{11}&=&E_{l=5}^{(0)}-\frac{3}{4}\langle\psi_{l=5}^{(0)}(r)| f(r)| \psi_{l=5}^{(0)}(r)\rangle ,\label{H61}\\
H_{12}&=&\sqrt{\frac{15}{2}}\langle\psi_{l=5}^{(0)}(r)|h_1(r)- h_2(r)| \psi_{l=5}^{(0)}(r)\rangle ,\label{H62}\\
H_{21}&=& H_{12}^*,\label{H63}\\
H_{22}&=& E_{l=5}^{(0)}+\langle\psi_{l=5}^{(0)}(r)| \frac{1}{4}f(r)+\frac{1}{2}g(r)\nonumber\\
&& -\frac{1}{2}h_1(r)-\frac{1}{2}h_2(r)\psi_{l=5}^{(0)}(r))\rangle .\label{H64}
\end{eqnarray}
With eqs.(\ref{H61}) $\sim$ (\ref{H64}), the cases with more $|^1H_5\rangle$ and $|^3H_5\rangle$ mixing states
can be obtained.

(7) The mass matrix of state $J^P=5^-$

The $J^P=5^-$ state is mixture of $^3G_5$ and $^3I_5$ states, both states with $s=1$, $l=4$ and $s=1$, $l=6$ can
construct the $J^P=5^-$ state. The basis for the mixing is $|\psi_1\rangle =|^3G_5\rangle$ and $|\psi_2\rangle =|^3I_5\rangle$.
The matrix elements of the mass matrix are
\begin{eqnarray}
H_{11}&=&E_{l=4}^{(0)}+\langle\psi_{l=4}^{(0)}(r)|[ \frac{1}{4}f(r)-\frac{2}{11}g(r)\nonumber\\
         &&+2h_1(r)+2h_2(r)]| \psi_{l=4}^{(0)}(r)\rangle ,\label{H65}\\
H_{12}&=&\frac{3}{11}\sqrt{\frac{15}{2}}\langle\psi_{l=4}^{(0)}(r)|g(r)| \psi_{l=6}^{(0)}(r)\rangle,\label{H66}\\
H_{21}&=& H_{12}^*,\label{H67}\\
H_{22}&=& E_{l=6}^{(0)}+\langle\psi_{l=6}^{(0)}(r)| [\frac{1}{4}f(r) -\frac{7}{22}g(r)\nonumber\\
    &&-\frac{7}{2} h_1(r)-\frac{7}{2} h_2(r)]| \psi_{l=6}^{(0)}(r)\rangle .\label{H68}
\end{eqnarray}
With eqs.(\ref{H65}) $\sim$ (\ref{H68}), the cases with more $|^3G_5\rangle$ and $|^3I_5\rangle$ mixing states
can be obtained.

In the above equations, the functions $f(r)$, $g(r)$, $h_1(r)$ and $h_2(r)$ are defined as
\begin{eqnarray}
f(r)&=&\frac{32\pi}{9m_1\tilde{m}_{2a}}\alpha_s(r)\delta_\sigma(r),\\
g(r)&=&\frac{4}{3}\frac{\alpha_s(r)}{m_1\tilde{m}_{2b}}\frac{1}{r^3},\\
h_1(r)&=&\left[\frac{4}{3}\frac{\alpha_s(r)}{r^3}\left( \frac{1}{m_1}+\frac{1}{\tilde{m}_{2c}}\right)\right.\nonumber\\
         &&\left.-\frac{1}{2r}\frac{\partial V(r)}{\partial r}\frac{1}{m_1}\right]\frac{1}{m_1},\\
h_2(r)&=&\left[\frac{4}{3}\frac{\alpha_s(r)}{r^3}\left( \frac{1}{m_1}+\frac{1}{\tilde{m}_{2c}}\right)\frac{1}{\tilde{m}_{2c}}\right.\nonumber\\
         &&\left.-\frac{1}{2r}\frac{\partial V(r)}{\partial r}\frac{1}{(\tilde{m}_{2d})^2}\right].
\end{eqnarray}

\section*{III Numerical result and discussion}
The parameters used in this work are the quark masses, the potential parameters $b$, $c$, $\tilde{m}_{2i}$, $\sigma_0$ and $s_0$. Comparing with our previous work \cite{LY}, we have slightly adjusted some of the parameters. The ground states are not sensitive to the value of the parameter $b$, which is slightly increased to give the correct predictions of excited states. The values we obtain by fitting are
\begin{eqnarray}
&& m_b=4.99\; {\rm GeV},\;\;\; m_c=1.58\; {\rm GeV},\nonumber\\
&&m_s=0.32\;{\rm GeV},\;\;\;  m_u=m_d=0.06\; {\rm GeV},\nonumber\\
&& b=0.175\;{\rm GeV}^2,\;\;\; c=-0.312\; {\rm GeV},\nonumber\\
&&\sigma_0=1.80\; {\rm GeV},\;\;\;s_0=1.55.
\end{eqnarray}
The values of $\tilde{m}_{2a}$, $\tilde{m}_{2b}$, $\tilde{m}_{2c}$ and $\tilde{m}_{2d}$ depend on the quark-antiquark system, they can
be written as
\begin{equation}
\tilde{m}_{2i}=\epsilon_i \tilde{m}_2,\; i=a,b,c,d.
\end{equation}
We find the values of $\epsilon_i$'s and $\tilde{m}_2$ are
\begin{equation}
(\epsilon_a,\epsilon_b,\epsilon_c,\epsilon_d)=(1.10,1.30,1.30,1.32)
\end{equation}
for $(b\bar{q})$ and $(c\bar{q})$ systems, and
\begin{equation}
(\epsilon_a,\epsilon_b,\epsilon_c,\epsilon_d)=(1.10,1.10,1.10,1.31)
\end{equation}
for $(b\bar{s})$ and $(c\bar{s})$ systems, and
\begin{equation}
\tilde{m}_2=\left\{\begin{array}{ll}
 0.562\;{\rm GeV}& {\rm for}\; (b\bar{q})\; {\rm system}, \\
 0.679\;{\rm GeV}& {\rm for}\; (b\bar{s})\; {\rm system},\\
 0.412\;{\rm GeV}& {\rm for}\; (c\bar{q})\; {\rm system},\\
 0.488\;{\rm GeV}& {\rm for}\; (c\bar{s})\; {\rm system},
 \end{array}\right.
\end{equation}
here $q$ is the light quark $u$ or $d$.

Numerical calculation shows that the solution is stable when $L>5\;{\rm fm}$, $N>50$. Here we take $L=10$ fm, $N=100$. The definition of $L$ and $N$ can be found in Refs. \cite{ymz,LY}.

The radial wave functions $\psi_{nl}(r)$ for $D$ meson in coordinate space are depicted in Fig.1 as an example. $|\psi_{nl}(r)|^2r^2$ is the possibility density distributed along the quark-antiquark distance $r$. We can use the Fourier transformation of the wave function in the coordinate space to get the momentum distribution, which is useful for studying $B$ and $D$ decays.

\begin{figure}[h]
\centering
\scalebox{0.8}{\epsfig{file=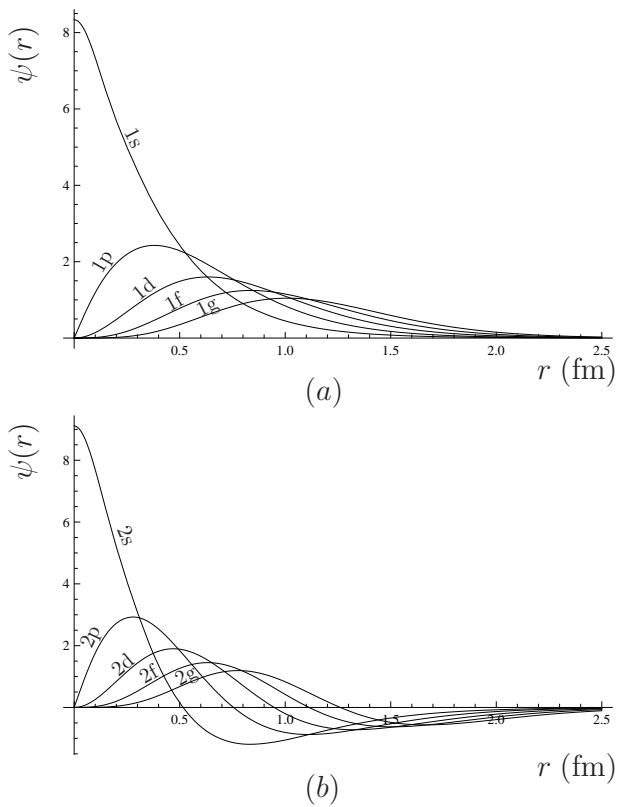}}
\caption{The radial wave function $\psi_{nl}(r)$ of $D$ meson in coordinate space, $l$ from 0 to 4. (a) is for $n=1$, (b) is for $n=2$.
}
\label{f1}
\end{figure}

In this work we calculate $B$, $D$, $B_s$, $D_s$ bound states with higher orbital quantum numbers, the quantum number $J^P$ of the new corresponding states are $2^-$, $3^-$, $3^+$, $4^+$, $4^-$. the numerical results for $B$, $D$, $B_s$, $D_s$ bound states are presented up to $l=4$ in four tables. The numerical results for $D$, $D_s$ bound states are given in Table I and II, the numerical results for $B$, $B_s$ bound states are given in Table III and IV. For completeness, results for ground-states and lower orbital excitations are also given. The mass predictions for lower orbital excitations have been calculated in Ref. \cite{GI} several decades before, which are also presented in Tables I to IV for comparison. The masses for most of the lower excitation states obtained in this work are approximately consistent with the masses given in Ref. \cite{GI}. Only for a few mesons, such as $D_1(2420)$, $D^*{s0}(2317)$, $D^*{s1}(2460)$ etc., the mass predictions are improved, the results in this work are more consistent with experiment. Mixings between states with appropriate quantum numbers are considered in our calculation. For doublets with $J^P=1^+$, $2^-$, $3^+$, $4^-$, mixing occurs more strongly. The mixing from state of higher radial quantum number with $n=2$ can not be neglected, it can be as large as 10\%. The masses measured in experiments are given in the last column in Tables I and III, from which we can see that the theoretical calculation in this work can accommodate the experimental data well.

Many new resonances with open charm flavor were observed in experiments recently \cite{D2010,LHCb}. Our mass predictions for these resonances are consistent with the experimental data. The masses measured by the LHCb collaboration are listed in the following \cite{LHCb}:
\begin{eqnarray}
 M(D_1(2420)^0)&=&2419.6\pm0.1\pm0.7\;\mbox{MeV}\nonumber\\
 M(D_2^*(2460)^0)&=&2460.4\pm0.4\pm1.2\;\mbox{MeV}\nonumber\\
 M(D_J^*(2650)^0)&=&2649.2\pm3.5\pm3.5\;\mbox{MeV}\nonumber\\
 M(D_J^*(2760)^0)&=&2761.1\pm5.1\pm6.5\;\mbox{MeV}\nonumber\\
 M(D_J(2580)^0)&=&2579.5\pm3.4\pm5.5\;\mbox{MeV}\nonumber\\
 M(D_J(2740)^0)&=&2737.0\pm3.5\pm11.2\;\mbox{MeV}\nonumber\\
 M(D_J(3000)^0)&=&2971.8\pm8.7\;\mbox{MeV}\nonumber\\
  M(D_2^*(2460)^0)&=&2460.4\pm0.1\pm0.1\;\mbox{MeV}\nonumber\\
  M(D_J^*(2760)^0)&=&2760.1\pm1.1\pm0.1\;\mbox{MeV}\nonumber\\
  M(D_J^*(3000)^0)&=&3008.1\pm4.0\;\mbox{MeV}\nonumber\\
     M(D_2^*(2460)^+)&=&2463.1\pm0.2\pm0.6\;\mbox{MeV}\nonumber\\
     M(D_J^*(2760)^+)&=&2771.7\pm1.7\pm3.8\;\mbox{MeV}\nonumber\\
     M(D_J^*(3000)^+)&=&3008.1\;\mbox{MeV}.\nonumber
 \end{eqnarray}
The assignments of the above observed states are listed in Table \ref{t1} and \ref{t2}. $D_J(2740)$ can be identified as mixing state of $|1^1D_2\rangle$ and $|1^3D_2\rangle$ with $J^P=2^-$. $D_J^*(2760)$ can be identified as dominantly $|1^3D_3\rangle$ state with $J^P=3^-$, which agrees with Ref. \cite{CFGN}. $D_J^*(3000)$ can be identified as mixing state of $|1^1F_3\rangle$ and $|1^3F_3\rangle$ with $J^P=3^+$. The other three observed states are $D_J(2580)$, $D_J^*(2650)$ and $D_J(3000)$, are identified as $n=2$ excited states in Table \ref{t2}. In Ref. \cite{LHCb}, $D_J(2580)$ is identified as $0^-$, $D_J^*(2650)$ is identified as $1^-$, but one can see that $D_J(2580)$ is not fully compatible with the assignment $0^-$ in our calculation. In this work, we take a different assignment, i.e., $D_J(2580)$ is identified as dominantly $|2^3S_1\rangle$ state with $J^P=1^-$, $D_J^*(2650)$ is identified as $|2^3P_0\rangle$ state with $J^P=0^+$, since our prediction for $|2^1S_0\rangle$ and $|2^3S_1\rangle$ state are lower than the predictions of Godfrey and Isgur \cite{GI}. The $D_J(3000)$ resonance is identified as dominantly $|2^3P_2\rangle$ state with $J^P=2^+$ in this work.

Since 2004 many new $D_s$ mesons have been observed, such as $D_{sJ}(2632)$ \cite{Ds2632}, $D_{sJ}^*(2860)$ \cite{Ds2860}, $D_{s1}^*(2710)$ \cite{Ds2700}, $D_{sJ}(3040)$ \cite{Ds2009}. One can see that our predicted mass spectrum for $D_s$ mesons below 2600 MeV accommodates the experimental data successfully \cite{PDG}, here we discuss the assignment of the recently observed $D_s$ meson resonances. The $D_{sJ}^*(2860)$ resonance parameters measured in Ref. \cite{Ds2009} are
\begin{eqnarray}
 M(D_{sJ}^*(2860)^+)&=&2862\pm2_{-2}^{+5}\;\mbox{MeV}\nonumber\\
 \Gamma&=&48\pm3\pm6\;\mbox{MeV}\nonumber
\end{eqnarray}
The $J^P=0^+$ assignment is proposed in Ref. \cite{BR}, but the observation of $D_{sJ}^*(2860)^+\to D^*K$ decay rules out the $J^P=0^+$ assignment \cite{Ds2009}. Our mass prediction for $|1^3D_3\rangle$ state is 2853 MeV, very close to the mass of $D_{sJ}^*(2860)$, thus we suggest that $D_{sJ}^*(2860)$ is $J^P=3^-$ state, which is also proposed in Ref. \cite{CFN}. For the $D_{s1}^*(2710)$ state, it is measured to be $2710\pm2_{-7}^{+12}$ MeV in Ref. \cite{LHCb}, the $J^P=1^-$ assignment is proposed in Refs. \cite{CFNR,CTLS}, comparing the experimental mass of $D_{s1}^*(2710)$ with our calculation for $|2^3S_1\rangle$ state in Table \ref{t2}, our prediction favors this assignment. The meson $D_{sJ}(2632)$ was firstly observed in Ref. \cite{Ds2632} at a mass of $2632.5\pm1.7$ MeV, in our predicted mass spectrum for $D_s$ mesons we assign it as the $|2^3P_0\rangle$ state with $J^P=0^-$, it is the closest states to 2632 MeV, although 50 MeV lower. The $D_{sJ}(3040)$ resonance is observed in the $D^*K$ mass spectrum at a mass of $3044\pm8_{-5}^{+30}$ in Ref. \cite{Ds2009}. The assignment of  this resonance is discussed in detail in Ref. \cite{CF1}. We identify it as $|2^1P_1\rangle$ with $J^P=1^+$, since its predicted mass 3012 MeV is the closest to the mass of the $D_{sJ}(3040)$ resonance.

For b-flavored meson sector, one can see the predicted mass spectra in Table \ref{t3} and \ref{t4} are in good agreement with experimental results. Although experimental information for b-flavored meson spectrum is limited, the measured $J^P=0^-$, $1^-$, $1^+$ and $2^+$ states can be an extra test to our model besides the charmed meson sector.

Next we would like to discuss the properties of the heavy-light quark-antiquark bound states from the point view of heavy quark symmetry. As discussed in Refs. \cite{CFGN,LY}, the spin of the heavy qurk decouples from the light degrees of freedom in the heavy quark limit. The total angular momentum of the light quark is $\vec{s}_l=\vec{s}_q +\vec{l}$, where $\vec{s}_q$ is the spin of the light quark, $\vec{l}$ the orbital angular momentum of the light quark. The total angular momentum of the heavy light meson is $\vec{J}=\vec{s}_l+\vec{s}_Q$, where $\vec{s}_Q$ is the spin of the heavy quark. Since the properties of the hadronic states do not depend on the spin $\vec{s}_Q$ and the flavor of the heavy quark due to the heavy quark symmetry, the heavy-light meson states only depend on the quantum numbers $s_l$ and $l$, thus we can classify the heavy-light mesons by these two quantum numbers. Heavy-light meson states can be divided into different doublets according to this classification. A doublet with $J^P_{s_l}=(0^-,1^-)_{1/2}$ exists for the case $l=0$, for example. Mesons within the same doublet degenerate in the heavy quark limit.  Since the parity of the meson is $P=(-1)^{l+1}$, the parity $P$ can denote the orbital angular momentum $l$ of the meson in a doublet. For each meson state with specific quantum numbers classified in the heavy quark limit, we can denote them by $|J^P_{s_l}\rangle$.

By analyzing angular momentum addition with the help of Clebsch-Gordan coefficients, we can decompose the basis states $|J^P_{s_l}\rangle$ in the heavy quark symmetry into combination of states with definite quantum numbers $l$, $S$ and $J$, i.e. the state $|^{2S+1}L_J\rangle$. The decompositions of all the doublets involved in this work are as follows:

(1) For $l=0$  states in the doublet $J^P_{s_l}=(0^-,1^-)_{1/2}$, we get
\begin{eqnarray}
|0^-_{1/2}\rangle &=&|^1S_0\rangle ,\label{e50}\\
|1^-_{1/2}\rangle &=&|^3S_1\rangle .\label{e51}
\end{eqnarray}

(2) For $l=1$  states in the doublets $J^P_{s_l}=(0^+,1^+)_{1/2}$ and $J^P_{s_l}=(1^+,2^+)_{3/2}$, we get
\begin{eqnarray}
|0^+_{1/2}\rangle &=&|^3P_0\rangle ,\label{e52}\\
|1^+_{1/2}\rangle &=&\sqrt{\frac{1}{3}}|^1P_1\rangle +\sqrt{\frac{2}{3}}|^3P_1\rangle,\\
|1^+_{3/2}\rangle &=&-\sqrt{\frac{2}{3}}|^1P_1\rangle +\sqrt{\frac{1}{3}}|^3P_1\rangle,\\
|2^+_{3/2}\rangle &=&|^3P_2\rangle .\label{e53}
\end{eqnarray}

(3) For $l=2$  states in the doublets $J^P_{s_l}=(1^-,2^-)_{3/2}$ and $J^P_{s_l}=(2^-,3^-)_{5/2}$, we get
\begin{eqnarray}
|1^-_{3/2}\rangle &=&|^3D_1\rangle ,\label{e54}\\
|2^-_{3/2}\rangle &=&\sqrt{\frac{2}{5}}|^1D_2\rangle +\sqrt{\frac{3}{5}}|^3D_2\rangle,\\
|2^-_{5/2}\rangle &=&-\sqrt{\frac{3}{5}}|^1D_2\rangle +\sqrt{\frac{2}{5}}|^3D_2\rangle,\\
|3^-_{5/2}\rangle &=&|^3D_3\rangle .\label{e55}
\end{eqnarray}

(4) For $l=3$  states in the doublets $J^P_{s_l}=(2^+,3^+)_{5/2}$ and $J^P_{s_l}=(3^+,4^+)_{7/2}$, we get
\begin{eqnarray}
|2^+_{5/2}\rangle &=&|^3F_2\rangle ,\label{e56}\\
|3^+_{5/2}\rangle &=&\sqrt{\frac{3}{7}}|^1F_3\rangle +2\sqrt{\frac{1}{7}}|^3F_3\rangle,\\
|3^+_{7/2}\rangle &=&-2\sqrt{\frac{1}{7}}|^1F_3\rangle +\sqrt{\frac{3}{7}}|^3F_3\rangle,\\
|4^+_{7/2}\rangle &=&|^3F_4\rangle .\label{e57}
\end{eqnarray}

(5) For $l=4$  states in the doublets $J^P_{s_l}=(3^-,4^-)_{7/2}$ and $J^P_{s_l}=(4^-,5^-)_{9/2}$, we get
\begin{eqnarray}
|3^-_{7/2}\rangle &=&|^3G_3\rangle ,\label{e58}\\
|4^-_{7/2}\rangle &=&\frac{2}{3}|^1G_4\rangle +\frac{\sqrt{5}}{3}|^3G_4\rangle,\\
|4^-_{9/2}\rangle &=&-\frac{\sqrt{5}}{3}|^1G_4\rangle +\frac{2}{3}|^3G_4\rangle,\\
|5^-_{9/2}\rangle &=&|^3G_5\rangle .\label{e59}
\end{eqnarray}
Particularly, for $P=(-1)^{J+1}$  unnatural parity states , they satisfy the relations
\begin{eqnarray}
|l^P_{l-1/2}\rangle &=&\sqrt{\frac{l}{2l+1}}|^1l_l\rangle +\sqrt{\frac{l+1}{2l+1}}|^3l_l\rangle,\\
|l^P_{l+1/2}\rangle &=&-\sqrt{\frac{l+1}{2l+1}}|^1l_l\rangle +\sqrt{\frac{l}{2l+1}}|^3l_l\rangle.
\end{eqnarray}

Consequently, the physical states with unnatural parity are linear combinations of $|^1l_l\rangle $ and $|^3l_l\rangle$, the mixing can be described by a mixing angle. We must note here that the sign of the mixing angle is fraught with ambiguities \cite{Godfrey,BBP}. It depends on the coupling order of the angular momentum and spins, $\vec{l}$, $\vec{s}_1$ and $\vec{s}_2$. The sign of the mixing angle will flip if the coupling order of the quark spins changes, for example.

In this work, the properties of the heavy flavored mesons implied by the heavy quark limit are approximately maintained in the calculation in the relativistic potential model. The mixing angle in the heavy quark limit should be $-\mbox{ArcSin}\sqrt{l/2l+1}$. For $l=1$, we get $-\mbox{ArcSin}\sqrt{1/3}=-0.615\; \mbox{rad}$, for $l=2$ we have $-\mbox{ArcSin}\sqrt{2/5}=-0.685\; \mbox{rad}$, etc. The mixing angle increases as $l$ increases. Comparing with the column ``Multiplet" in the Table I$\sim$IV for each meson, one can find that this property is maintained in the numerical results. As discussed in our previous work \cite{LY}, only the mixing between $|1^1P_1\rangle$ and $|1^3P_1\rangle$ in $1^+$ state of $c\bar{q}$ is very small, which seriously deviates from the result in the heavy quark limit $-0.615\; \mbox{rad}$, the reason has been explained in our previous work \cite{LY}.  The Belle collaboration determined this mixing angle,which is $\theta =-0.10\pm 0.03\pm 0.02\pm 0.02\; \mbox{rad}$ \cite{Belle}. Our prediction with small mixing angle for $1^+$ state of $c\bar{q}$ is well consistent with Belle's measurement.

Finally the wave function of each bound state can be obtained simultaneously when solving the wave equation. Figure \ref{f1} is the radial wave functions for $D$ meson as an example. It is easy to get all the wave functions when they are needed.

\section*{IV Summary}
The bound states of heavy-light quark and antiquark system are studied in the relativistic potential model. The predictions of heavy-light charmed and $b$-flavored meson states are presented. The low-lying meson states are in good agreement with experiments. The higher excitation states are particularly focused on. All the observed mesons can be successfully accommodated in our calculation. The wave functions of each bound state can be obtained by solving the wave equation and be used in the study of $B$ and $D$ decays.

\newpage

\begin{widetext}
\begin{center}
\begin{table}
\caption{Theoretical spectrum of  $(c\bar{q})$ and $(c\bar{s})$ bound states mainly with
the radial quantum number $n=1$. The results in the column labeled ``GI" are theoretical
masses from Ref. \cite{GI}.}
 \label{t1}
\begin{tabular}{|c|c|c|c|c|c|c|}\hline
& Meson & $J^P$ & Multiplet & Mass (GeV) & GI (GeV)& Exp. (MeV) \cite{LHCb,PDG} \\ \hline
 &$D$ & $0^-$ & $|1^1S_0\rangle$ & 1.867 & 1.88 & $1869.62\pm 0.15$ \\ \cline{2-7}
&$D^*$&$1^-$ & $\begin{array}{cc} &0.997|1^3S_1\rangle -0.035|1^3D_1\rangle +0.051|2^3S_1\rangle+0.033|2^3D_1\rangle\\
     &-0.026|3^3S_1\rangle-0.031|3^3D_1\rangle\end{array}$& 2.004 & 2.04 & $2010.28\pm 0.13$ \\ \cline{4-7}
&     &      & $\begin{array}{cc} &0.046|1^3S_1\rangle+0.979|1^3D_1\rangle-0.090|2^3S_1\rangle-0.148|2^3D_1\rangle \\
     &-0.041|3^3S_1\rangle+0.093|3^3D_1\rangle\end{array}$& 2.785 &2.82 &  \\ \cline{2-7}
& $D_0^*(2400)^0$  & $0^+$ & $|1^3P_0\rangle$ & 2.302& 2.40 & $2318\pm 29$ \\ \cline{2-7}
&                & $1^+$ &$0.017|1^1P_1\rangle +0.990|1^3P_1\rangle -0.076|2^1P_1\rangle -0.117|2^3P_1\rangle$  & 2.415& 2.44 &     \\ \cline{4-6}
& $D_1(2420)$  &       & $0.996|1^1P_1\rangle -0.028|1^3P_1\rangle-0.028|2^1P_1\rangle -0.076|2^3P_1\rangle$ & 2.429 & 2.49& $2421.3\pm 0.6$ \\ \cline{2-7}
&$D_2^*(2460)$ & $2^+$ &  $\begin{array}{cc}&0.988|1^3P_2\rangle -0.014|1^3F_2\rangle+0.135|2^3P_2\rangle +0.011|2^3F_2\rangle\\
            &-0.071|3^3P_2\rangle -0.009|3^3F_2\rangle\end{array}$ &2.468& 2.50 & $2464.4\pm 1.9$ \\ \cline{2-2}\cline{4-7}
&              &       &  $\begin{array}{cc}&0.017|1^3P_2\rangle +0.997|1^3F_2\rangle-0.024|2^3P_2\rangle -0.047|2^3F_2\rangle\\
            &-0.021|3^3P_2\rangle +0.039|3^3F_2\rangle\end{array}$ & 3.123 & &             \\
            \cline{2-7}
&$D_J(2740)^0$  & $2^-$ & $\begin{array}{cc}&0.769|1^1D_2\rangle -0.639|1^3D_2\rangle+0.0003|2^1D_2\rangle +0.007|2^3D_2\rangle\\
            &-0.007|3^1D_2\rangle +0.002|3^3D_2\rangle\end{array}$ &2.737 && $2737.0\pm 3.5\pm11.2$ \\ \cline{2-2}\cline{4-7}
&              &       & $\begin{array}{cc}&0.639|1^1D_2\rangle +0.768|1^3D_2\rangle-0.005|2^1D_2\rangle -0.021|2^3D_2\rangle\\
            &0.013|3^1D_2\rangle +0.022|3^3D_2\rangle\end{array} $& 2.834 & &            \\
            \cline{2-7}
$(c\bar{q})$&$D_J^*(2760)^0$& $3^-$ & $\begin{array}{cc}&0.997|1^3D_3\rangle -0.008|1^3G_3\rangle+0.063|2^3D_3\rangle +0.006|2^3G_3\rangle\\
            &-0.039|3^3D_3\rangle +0.005|3^3G_3\rangle\end{array}$ & 2.754 && $2760.1\pm 1.1\pm3.7$\\ \cline{2-2}\cline{4-7}
&              &       & $\begin{array}{cc}&0.008|1^3D_3\rangle +0.9997|1^3G_3\rangle-0.007|2^3D_3\rangle +0.002|2^3G_3\rangle\\
            &-0.009|3^3D_3\rangle +0.018|3^3G_3\rangle\end{array}$  & 3.397 & &            \\
            \cline{2-7}
&$D^*_J(3000)^0$& $3^+$ & $\begin{array}{cc}&0.759|1^1F_3\rangle -0.650|1^3F_3\rangle-0.016|2^1F_3\rangle +0.017|2^3F_3\rangle\\
            &-0.001|3^1F_3\rangle -0.001|3^3F_3\rangle\end{array}$ & 2.997 &  & $3008.1\pm 4.0$  \\ \cline{2-2}\cline{4-7}
&              &       & $\begin{array}{cc}&0.650|1^1F_3\rangle +0.759|1^3F_3\rangle+0.020|2^1F_3\rangle +0.016|2^3F_3\rangle\\
            &+0.002|3^1F_3\rangle +0.006|3^3F_3\rangle\end{array}$  & 3.138 & &            \\
            \cline{2-7}
& & $4^+$ & $\begin{array}{cc}&0.9995|1^3F_4\rangle -0.005|1^3H_4\rangle+0.021|2^3F_4\rangle +0.004|2^3H_4\rangle\\
            &-0.022|3^3F_4\rangle -0.003|3^3H_4\rangle\end{array}$ & 3.003 &  &   \\ \cline{2-2}\cline{4-7}
&              &       & $\begin{array}{cc}&0.005|1^3F_4\rangle +0.9994|1^3H_4\rangle-0.003|2^3F_4\rangle +0.031|2^3H_4\rangle\\
            &-0.007|3^3F_4\rangle +0.008|3^3H_4\rangle\end{array}$  &3.638 & &            \\
            \cline{2-7}
& & $4^-$ & $\begin{array}{cc}&0.750|1^1G_4\rangle -0.660|1^3G_4\rangle-0.028|2^1G_4\rangle +0.026|2^3G_4\rangle\\
            &+0.003|3^1G_4\rangle -0.003|3^3G_4\rangle\end{array}$ & 3.229 &  &   \\ \cline{2-2}\cline{4-7}
&              &       & $\begin{array}{cc}&0.660|1^1G_4\rangle +0.750|1^3G_4\rangle+0.034|2^1G_4\rangle +0.032|2^3G_4\rangle\\
            &-0.002|3^1G_4\rangle +0.0002|3^3G_4\rangle\end{array}$  &3.398 & &            \\\hline
 &$D_s^\pm$ & $0^-$ & $|1^1S_0\rangle$ & 1.963 & 1.98 & $1968.49\pm 0.32$ \\ \cline{2-7}
&$D_s^{*\pm}$&$1^-$ & $\begin{array}{cc} &0.996|1^3S_1\rangle -0.045|1^3D_1\rangle +0.052|2^3S_1\rangle+0.041|2^3D_1\rangle\\
     &-0.028|3^3S_1\rangle-0.036|3^3D_1\rangle\end{array}$ & 2.103 & 2.13 & $2112.3\pm 0.5$ \\ \cline{2-2}\cline{4-7}
&     &      & $\begin{array}{cc}&0.064|1^3S_1\rangle +0.950|1^3D_1\rangle -0.191|2^3S_1\rangle-0.201|2^3D_1\rangle\\
     &-0.047|3^3S_1\rangle+0.117|3^3D_1\rangle\end{array}$& 2.803& 2.90 &  \\ \cline{2-7}
& $D_{s0}^*(2317)^0$  & $0^+$ & $|1^3P_0\rangle$ & 2.317 & 2.48 & $2317.8\pm 0.6$ \\ \cline{2-7}
& $D_{s1}(2460)$ & $1^+$ & $0.475|1^1P_1\rangle +0.851|1^3P_1\rangle -0.117|2^1P_1\rangle -0.193|2^3P_1\rangle$  & 2.436 &2.53 &  $2459.6\pm 0.6$ \\ \cline{2-2}\cline{4-7}
& $D_{s1}(2536)$ &       & $0.871|1^1P_1\rangle -0.488|1^3P_1\rangle+0.040|2^1P_1\rangle-0.033|2^3P_1\rangle$&2.533 & 2.57 &  $2535.12\pm 0.13$ \\ \cline{2-7}
&$D_{s2}^*(2573)$ & $2^+$ &$\begin{array}{cc}&0.983|1^3P_2\rangle -0.018|1^3F_2\rangle+0.164|2^3P_2\rangle +0.013|2^3F_2\rangle\\
            &-0.080|3^3P_2\rangle -0.010|3^3F_2\rangle\end{array}$ & 2.573&2.59 & $2571.9\pm 0.8$ \\ \cline{2-2}\cline{4-7}
&              &       & $\begin{array}{cc}&0.027|1^3P_2\rangle +0.991|1^3F_2\rangle-0.063|2^3P_2\rangle -0.097|2^3F_2\rangle\\
            &-0.036|3^3P_2\rangle +0.058|3^3F_2\rangle\end{array}$ & 3.145& &             \\
            \cline{2-7}
& & $2^-$ & $\begin{array}{cc}&0.740|1^1D_2\rangle -0.672|1^3D_2\rangle+0.027|2^1D_2\rangle -0.007|2^3D_2\rangle\\
            &-0.017|3^1D_2\rangle +0.005|3^3D_2\rangle\end{array}$ &2.835 &&  \\ \cline{2-2}\cline{4-7}
&              &       & $\begin{array}{cc}&0.671|1^1D_2\rangle +0.737|1^3D_2\rangle-0.037|2^1D_2\rangle -0.063|2^3D_2\rangle\\
            &+0.025|3^1D_2\rangle +0.039|3^3D_2\rangle\end{array} $& 2.864 & &            \\
            \cline{2-7}
$(c\bar{s})$&$D_{sJ}^*(2860)$& $3^-$ & $\begin{array}{cc}&0.994|1^3D_3\rangle -0.009|1^3G_3\rangle+0.093|2^3D_3\rangle +0.007|2^3G_3\rangle\\
            &-0.047|3^3D_3\rangle -0.005|3^3G_3\rangle\end{array}$ & 2.853 &&$2862\pm2_{-2}^{+5}$ \cite{Ds2009} \\ \cline{2-2}\cline{4-7}
&              &       & $\begin{array}{cc}&0.011|1^3D_3\rangle +0.998|1^3G_3\rangle-0.014|2^3D_3\rangle -0.043|2^3G_3\rangle\\
            &-0.010|3^3D_3\rangle +0.031|3^3G_3\rangle\end{array}$  & 3.416 & &            \\
            \cline{2-7}
&$ $ & $3^+$ & $\begin{array}{cc}&0.762|1^1F_3\rangle -0.648|1^3F_3\rangle+0.006|2^1F_3\rangle -0.002|2^3F_3\rangle\\
            &-0.008|3^1F_3\rangle +0.005|3^3F_3\rangle\end{array}$ & 3.092 &  &   \\ \cline{2-2}\cline{4-7}
&              &       & $\begin{array}{cc}&0.648|1^1F_3\rangle +0.761|1^3F_3\rangle-0.008|2^1F_3\rangle -0.016|2^3F_3\rangle\\
            &+0.010|3^1F_3\rangle +0.015|3^3F_3\rangle\end{array}$  & 3.165 & &            \\
            \cline{2-7}
& & $4^+$ & $\begin{array}{cc}&0.998|1^3F_4\rangle -0.006|1^3H_4\rangle+0.051|2^3F_4\rangle +0.004|2^3H_4\rangle\\
            &-0.030|3^3F_4\rangle-0.003|3^3H_4\rangle\end{array}$ & 3.097 &  &   \\ \cline{2-2}\cline{4-7}
&              &       & $\begin{array}{cc}&0.006|1^3F_4\rangle +0.9997|1^3H_4\rangle-0.006|2^3F_4\rangle -0.010|2^3H_4\rangle\\
            &-0.005|3^3F_4\rangle +0.018|3^3H_4\rangle\end{array}$  &3.653 & &            \\
                        \cline{2-7}
& & $4^-$ & $\begin{array}{cc}&0.754|1^1G_4\rangle -0.657|1^3G_4\rangle-0.005|2^1G_4\rangle +0.006|2^3G_4\rangle\\
            &-0.004|3^1G_4\rangle +0.002|3^3G_4\rangle\end{array}$ & 3.320 &  &   \\ \cline{2-2}\cline{4-7}
&              &       & $\begin{array}{cc}&0.657|1^1G_4\rangle +0.754|1^3G_4\rangle+0.007|2^1G_4\rangle +0.004|2^3G_4\rangle\\
            &+0.004|3^1G_4\rangle +0.007|3^3G_4\rangle\end{array}$  &3.420 & &            \\
\hline
\end{tabular}
\end{table}
\end{center}

\begin{center}
\begin{table}
\caption{Theoretical spectrum of  $(c\bar{q})$ and $(c\bar{s})$ bound states mainly with
the radial quantum number $n=2$.}
 \label{t2}
\begin{tabular}{|c|c|c|c|c|c|c|}\hline
     & $J^P$ & Multiplet & Mass (GeV) & GI (GeV) & Meson & Exp. (MeV) \\
     \hline
  & $0^-$ & $|2^1S_0\rangle$ & 2.483 &2.58& & \\ \cline{2-7}
 &$1^-$ & $\begin{array}{cc} &-0.043|1^3S_1\rangle+0.083|1^3D_1\rangle+0.991|2^3S_1\rangle-0.055|2^3D_1\rangle \\
     &+0.065|3^3S_1\rangle+0.067|3^3D_1\rangle\end{array}$& 2.613 &2.64&$D_J(2580)^0$ & $2579.5\pm 3.4\pm 5.5$\cite{LHCb}\\ \cline{3-7}
 &      & $\begin{array}{cc} &-0.040|1^3S_1\rangle+0.161|1^3D_1\rangle+0.058|2^3S_1\rangle+0.951|2^3D_1\rangle \\
     &-0.180|3^3S_1\rangle-0.180|3^3D_1\rangle\end{array}$& 3.155 && &\\ \cline{2-7}
 & $0^+$ & $|2^3P_0\rangle$ & 2.685 &&$D_J^*(2650)^0$ & $2649.2\pm 3.5\pm 3.5$  \cite{LHCb}\\ \cline{2-7}
  & $1^+$ &$0.083|1^1P_1\rangle +0.137|1^3P_1\rangle+0.4755|2^1P_1\rangle+0.867|2^3P_1\rangle$ & 2.865 & & & \\ \cline{3-7}
 &       & $-0.010|1^1P_1\rangle +0.014|1^3P_1\rangle+0.887|2^1P_1\rangle-0.462|2^3P_1\rangle$& 2.902 && & \\ \cline{2-7}
 & $2^+$ &  $\begin{array}{cc}& -0.111|1^3P_2\rangle +0.029|1^3F_2\rangle+0.959|2^3P_2\rangle -0.021|2^3F_2\rangle\\
            &+0.258|3^3P_2\rangle +0.016|3^3F_2\rangle\end{array}$ & 2.969 & & $D_J(3000)^0$ & $2971.8\pm 8.7$ \cite{LHCb}\\ \cline{3-7}
  &       &  $\begin{array}{cc}&-0.057|1^3P_2\rangle +0.040|1^3F_2\rangle+0.124|2^3P_2\rangle +0.898|2^3F_2\rangle\\
            &-0.412|3^3P_2\rangle -0.059|3^3F_2\rangle\end{array}$ & 3.458 && &  \\
            \cline{2-7}
 $(c\bar{q})$ & $2^-$ &  $\begin{array}{cc}&-0.003|1^1D_2\rangle -0.011|1^3D_2\rangle+0.757|2^1D_2\rangle -0.653|2^3D_2\rangle\\
            &+0.008|3^1D_2\rangle +0.005|3^3D_2\rangle\end{array}$ &3.144 && & \\ \cline{3-7}
  &       &  $\begin{array}{cc}&0.009|1^1D_2\rangle +0.020|1^3D_2\rangle+0.653|2^1D_2\rangle +0.755|2^3D_2\rangle\\
            &-0.021|3^1D_2\rangle -0.044|3^3D_2\rangle\end{array}$ & 3.211 &&  & \\
            \cline{2-7}
 & $3^-$ &  $\begin{array}{cc}&-0.057|1^3D_3\rangle +0.009|1^3G_3\rangle+0.990|2^3D_3\rangle -0.010|2^3G_3\rangle\\
            &+0.124|3^3D_3\rangle +0.008|3^3G_3\rangle\end{array}$ & 3.175 && & \\ \cline{3-7}
  &       &  $\begin{array}{cc}&-0.007|1^3D_3\rangle -0.002|1^3G_3\rangle+0.012|2^3D_3\rangle +0.9996|2^3G_3\rangle\\
            &-0.018|3^3D_3\rangle -0.015|3^3G_3\rangle\end{array}$ & 3.704 &&  & \\
            \cline{2-7}
 & $3^+$ &  $\begin{array}{cc}&0.145|1^1F_3\rangle -0.020|1^3F_3\rangle+0.755|2^1F_3\rangle -0.654|2^3F_3\rangle\\
            &-0.017|3^1F_3\rangle +0.020|3^3F_3\rangle\end{array}$ & 3.360 && & \\ \cline{3-7}
  &       &  $\begin{array}{cc}&-0.018|1^1F_3\rangle -0.018|1^3F_3\rangle+0.654|2^1F_3\rangle +0.755|2^3F_3\rangle\\
            &+0.020|3^1F_3\rangle +0.011|3^3F_3\rangle\end{array}$ & 3.480 && &  \\
            \cline{2-7}
 & $4^+$ &  $\begin{array}{cc}&-0.020|1^3F_4\rangle +0.004|1^3H_4\rangle+0.998|2^3F_4\rangle -0.006|2^3H_4\rangle\\
            &+0.050|3^3F_4\rangle +0.005|3^3H_4\rangle\end{array}$ & 3.373 && & \\ \cline{3-7}
  &       &  $\begin{array}{cc}&-0.004|1^3F_4\rangle -0.032|1^3H_4\rangle+0.006|2^3F_4\rangle +0.999|2^3H_4\rangle\\
            &-0.006|3^3F_4\rangle +0.029|3^3H_4\rangle\end{array}$ & 3.926 && &  \\

\hline
   & $0^-$ & $|2^1S_0\rangle$ & 2.582&2.67& $D_{sJ}(2632)$&$2632.5\pm 1.7$ \cite{Ds2632}\\ \cline{2-7}
 &$1^-$ & $\begin{array}{cc} & -0.035|1^3S_1\rangle+0.175|1^3D_1\rangle+0.976|2^3S_1\rangle-0.087|2^3D_1\rangle \\
     &+0.063|3^3S_1\rangle+0.065|3^3D_1\rangle\end{array}$ & 2.709 & 2.73&$D_{s1}^*(2710)$ & $2708\pm9^{+11}_{10}$\cite{Ds2700} \\ \cline{3-7}
  &      & $\begin{array}{cc} &-0.050|1^3S_1\rangle+0.190|1^3D_1\rangle+0.082|2^3S_1\rangle+0.816|2^3D_1\rangle \\
     &-0.507|3^3S_1\rangle-0.180|3^3D_1\rangle\end{array}$& 3.191 && & \\ \cline{2-7}
 & $0^+$ & $|2^3P_0\rangle$ & 2.707&& & \\ \cline{2-7}
 & $1^+$ & $0.117|1^1P_1\rangle +0.193|1^3P_1\rangle+0.498|2^1P_1\rangle+0.837|2^3P_1\rangle$ & 2.911 & & & \\ \cline{3-7}
 &       & $-0.044|1^1P_1\rangle +0.026|1^3P_1\rangle +0.858|2^1P_1\rangle -0.510|2^3P_1\rangle$ & 3.012 & &$D_{sJ}(3040)$ &$3044\pm8^{+30}_{-5}$\cite{Ds2009}\\ \cline{2-7}
  & $2^+$ &$\begin{array}{cc}& -0.130|1^3P_2\rangle +0.070|1^3F_2\rangle+0.941|2^3P_2\rangle -0.030|2^3F_2\rangle\\
            &+0.301|3^3P_2\rangle +0.022|3^3F_2\rangle\end{array}$ &3.070 && &  \\ \cline{3-7}
  &       & $\begin{array}{cc}&-0.001|1^3P_2\rangle +0.108|1^3F_2\rangle-0.011|2^3P_2\rangle +0.976|2^3F_2\rangle\\
            &+0.119|3^3P_2\rangle -0.143|3^3F_2\rangle\end{array}$ & 3.477 && & \\
             \cline{2-7}
  $(c\bar{s})$ & $2^-$ &$\begin{array}{cc}&0.048|1^1D_2\rangle +0.041|1^3D_2\rangle-0.087|2^1D_2\rangle +0.990|2^3D_2\rangle\\
            &-0.072|3^1D_2\rangle -0.051|3^3D_2\rangle\end{array}$ &3.240 && &  \\ \cline{3-7}
  &       & $\begin{array}{cc}&0.011|1^1D_2\rangle +0.051|1^3D_2\rangle+0.992|2^1D_2\rangle +0.080|2^3D_2\rangle\\
            &-0.009|3^1D_2\rangle -0.080|3^3D_2\rangle\end{array}$ & 3.246 & & &\\
              \cline{2-7}
 & $3^-$ &  $\begin{array}{cc}&-0.083|1^3D_3\rangle +0.015|1^3G_3\rangle+0.982|2^3D_3\rangle -0.013|2^3G_3\rangle\\
            &+0.171|3^3D_3\rangle +0.010|3^3G_3\rangle\end{array}$ & 3.272 &&&  \\ \cline{3-7}
  &       &  $\begin{array}{cc}&-0.011|1^3D_3\rangle +0.044|1^3G_3\rangle+0.021|2^3D_3\rangle +0.995|2^3G_3\rangle\\
            &-0.051|3^3D_3\rangle -0.071|3^3G_3\rangle\end{array}$ & 3.724 && &  \\
            \cline{2-7}
 & $3^+$ &  $\begin{array}{cc}&-0.007|1^1F_3\rangle +0.0002|1^3F_3\rangle+0.754|2^1F_3\rangle -0.657|2^3F_3\rangle\\
            &+0.016|3^1F_3\rangle -0.007|3^3F_3\rangle\end{array}$ & 3.454 &&&  \\ \cline{3-7}
  &       &  $\begin{array}{cc}&0.010|1^1F_3\rangle +0.016|1^3F_3\rangle+0.657|2^1F_3\rangle +0.753|2^3F_3\rangle\\
            &-0.019|3^1F_3\rangle -0.033|3^3F_3\rangle\end{array}$ & 3.507 & & & \\
            \cline{2-7}
 & $4^+$ &  $\begin{array}{cc}&-0.048|1^3F_4\rangle +0.006|1^3H_4\rangle+0.994|2^3F_4\rangle -0.007|2^3H_4\rangle\\
            &+0.096|3^3F_4\rangle+0.006|3^3H_4\rangle\end{array}$ & 3.467 &&&  \\ \cline{3-7}
  &       &  $\begin{array}{cc}&-0.005|1^3F_4\rangle +0.011|1^3H_4\rangle+0.008|2^3F_4\rangle +0.9994|2^3H_4\rangle\\
            &-0.012|3^3F_4\rangle -0.027|3^3H_4\rangle\end{array}$ & 3.941 && &  \\

\hline
\end{tabular}
\end{table}
\end{center}

\begin{center}
\begin{table}
\caption{Theoretical spectrum of  $(b\bar{q})$ and $(b\bar{s})$ bound states mainly with
the radial quantum number $n=1$.}
 \label{t3}
\begin{tabular}{|c|c|c|c|c|c|c|}\hline
& Meson & $J^P$ & Multiplet & Mass (GeV) & GI (GeV)& Exp. (MeV) \cite{PDG} \\ \hline
 &$B$ & $0^-$ & $|1^1S_0\rangle$ & 5.271 & 5.31& $5279.25\pm 0.17$ \\ \cline{2-7}
&$B^*$&$1^-$ & $\begin{array}{cc}&0.9996|1^3S_1\rangle -0.012|1^3D_1\rangle +0.019|2^3S_1\rangle +0.012|2^3D_1\rangle\\
      &-0.009|3^3S_1\rangle -0.011|3^3D_1\rangle\end{array}$
 & 5.317 & 5.32& $5325.2\pm 0.4$ \\ \cline{4-7}
&     &      & $\begin{array}{cc} &0.014|1^3S_1\rangle+0.995|1^3D_1\rangle-0.020|2^3S_1\rangle-0.080|2^3D_1\rangle \\
     &-0.011|3^3S_1\rangle+0.051|3^3D_1\rangle\end{array}$ & 6.087 & &  \\ \cline{2-7}
&     & $0^+$ & $|1^3P_0\rangle$ & 5.696 & &  \\ \cline{2-7}
&$B_1(5721)$ & $1^+$ & $0.511|1^1P_1\rangle +0.847|1^3P_1\rangle-0.082|2^1P_1\rangle-0.117|2^3P_1\rangle$ & 5.732 & & $5723.5\pm 2.0$ \\ \cline{4-6}
&            &       & $0.855|1^1P_1\rangle -0.517|1^3P_1\rangle+0.031|2^1P_1\rangle-0.031|2^3P_1\rangle$  & 5.759 & &                \\ \cline{2-7}
&$B_2^*(5747)$ & $2^+$ & $\begin{array}{cc}&0.995|1^3P_2\rangle -0.005|1^3F_2\rangle+0.090|2^3P_2\rangle -0.004|2^3F_2\rangle\\
            &-0.049|3^3P_2\rangle -0.004|3^3F_2\rangle\end{array}$ & 5.775 &5.8 & $5743\pm 5$ \\ \cline{2-2}\cline{4-7}
&              &       & $\begin{array}{cc}&0.006|1^3P_2\rangle +0.9995|1^3F_2\rangle-0.008|2^3P_2\rangle -0.021|2^3F_2\rangle\\
            &-0.006|3^3P_2\rangle +0.021|3^3F_2\rangle\end{array}$  & 6.374 & &            \\
            \cline{2-7}
&& $2^-$ & $\begin{array}{cc}&0.764|1^1D_2\rangle -0.645|1^3D_2\rangle+0.010|2^1D_2\rangle -0.005|2^3D_2\rangle\\
            &-0.010|3^1D_2\rangle +0.007|3^3D_2\rangle\end{array}$ &6.051 && \\ \cline{2-2}\cline{4-7}
&              &       & $\begin{array}{cc}&0.644|1^1D_2\rangle +0.763|1^3D_2\rangle-0.017|2^1D_2\rangle -0.025|2^3D_2\rangle\\
            &0.016|3^1D_2\rangle +0.021|3^3D_2\rangle\end{array} $&6.105 & &            \\
            \cline{2-7}
$(b\bar{q})$&& $3^-$ & $\begin{array}{cc}&0.999|1^3D_3\rangle -0.003|1^3G_3\rangle+0.038|2^3D_3\rangle +0.002|2^3G_3\rangle\\
            &-0.025|3^3D_3\rangle -0.002|3^3G_3\rangle\end{array}$ & 6.061 &&\\ \cline{2-2}\cline{4-7}
&              &       & $\begin{array}{cc}&0.003|1^3D_3\rangle +0.9999|1^3G_3\rangle-0.003|2^3D_3\rangle +0.007|2^3G_3\rangle\\
            &-0.002|3^3D_3\rangle +0.010|3^3G_3\rangle\end{array}$  & 6.614 & &            \\
            \cline{2-7}
&$ $ & $3^+$ & $\begin{array}{cc}&0.753|1^1F_3\rangle -0.658|1^3F_3\rangle-0.006|2^1F_3\rangle +0.007|2^3F_3\rangle\\
            &-0.004|3^1F_3\rangle +0.003|3^3F_3\rangle\end{array}$ & 6.296 &  &   \\ \cline{2-2}\cline{4-7}
&              &       & $\begin{array}{cc}&0.658|1^1F_3\rangle +0.753|1^3F_3\rangle+0.007|2^1F_3\rangle +0.005|2^3F_3\rangle\\
            &+0.005|3^1F_3\rangle +0.007|3^3F_3\rangle\end{array}$  & 6.383 & &            \\
            \cline{2-7}
& & $4^+$ & $\begin{array}{cc}&0.9999|1^3F_4\rangle -0.002|1^3H_4\rangle+0.011|2^3F_4\rangle +0.001|2^3H_4\rangle\\
            &-0.013|3^3F_4\rangle -0.001|3^3H_4\rangle\end{array}$ & 6.302 &  &   \\ \cline{2-2}\cline{4-7}
&              &       & $\begin{array}{cc}&0.002|1^3F_4\rangle +0.9997|1^3H_4\rangle-0.002|2^3F_4\rangle +0.024|2^3H_4\rangle\\
            &-0.001|3^3F_4\rangle +0.004|3^3H_4\rangle\end{array}$  &6.826 & &            \\
            \cline{2-7}
& & $4^-$ & $\begin{array}{cc}&0.744|1^1G_4\rangle -0.667|1^3G_4\rangle-0.017|2^1G_4\rangle +0.016|2^3G_4\rangle\\
            &-0.0004|3^1G_4\rangle -0.0001|3^3G_4\rangle\end{array}$ & 6.511 &  &   \\ \cline{2-2}\cline{4-7}
&              &       & $\begin{array}{cc}&0.667|1^1G_4\rangle +0.744|1^3G_4\rangle+0.020|2^1G_4\rangle +0.020|2^3G_4\rangle\\
            &+0.001|3^1G_4\rangle +0.002|3^3G_4\rangle\end{array}$  &6.619 & &            \\\hline
&$B_s$ & $0^-$ & $|1^1S_0\rangle$ & 5.360 & 5.39 & $5366.77\pm 0.24$ \\ \cline{2-7}
&$B_s^*$&$1^-$ &  $\begin{array}{cc} &0.9995|1^3S_1\rangle -0.014|1^3D_1\rangle +0.020|2^3S_1\rangle+0.013|2^3D_1\rangle\\
     &-0.010|3^3S_1\rangle-0.012|3^3D_1\rangle\end{array}$ & 5.408 & 5.45 & $5415.4^{+2.4}_{-2.1} $ \\ \cline{4-7}
&     &      &$\begin{array}{cc} &0.017|1^3S_1\rangle+0.989|1^3D_1\rangle-0.030|2^3S_1\rangle-0.122|2^3D_1\rangle \\
     &-0.010|3^3S_1\rangle-0.069|3^3D_1\rangle\end{array}$& 6.120 & &  \\ \cline{2-7}
&     & $0^+$ & $|1^3P_0\rangle$ & 5.731 & &  \\ \cline{2-7}
&             & $1^+$ &$0.537|1^1P_1\rangle +0.822|1^3P_1\rangle-0.108|2^1P_1\rangle-0.159|2^3P_1\rangle$ & 5.765 & &  \\ \cline{2-2}\cline{4-7}
&$B_{s1}(5830)$ &     &$0.835|1^1P_1\rangle -0.546|1^3P_1\rangle+0.056|2^1P_1\rangle-0.042|2^3P_1\rangle$ & 5.850& & $5829.4\pm 0.7$ \\ \cline{2-7}
&$B_{s2}^*(5840)$ & $2^+$ &  $\begin{array}{cc}&0.992|1^3P_2\rangle -0.006|1^3F_2\rangle+0.112|2^3P_2\rangle +0.005|2^3F_2\rangle\\
            &-0.056|3^3P_2\rangle -0.004|3^3F_2\rangle\end{array}$ & 5.866 & 5.88 & $5839.7\pm 0.6$ \\ \cline{2-2}\cline{4-7}
&              &       &  $\begin{array}{cc}&0.008|1^3P_2\rangle +0.998|1^3F_2\rangle-0.015|2^3P_2\rangle -0.056|2^3F_2\rangle\\
            &-0.008|3^3P_2\rangle +0.033|3^3F_2\rangle\end{array}$& 6.405&  &          \\
            \cline{2-7}
& & $2^-$ & $\begin{array}{cc}&-0.684|1^1D_2\rangle +0.728|1^3D_2\rangle-0.033|2^1D_2\rangle +0.011|2^3D_2\rangle\\
            &+0.019|3^1D_2\rangle -0.007|3^3D_2\rangle\end{array}$ &6.134 &&  \\ \cline{2-2}\cline{4-7}
&              &       & $\begin{array}{cc}&0.727|1^1D_2\rangle +0.682|1^3D_2\rangle-0.037|2^1D_2\rangle -0.058|2^3D_2\rangle\\
            &+0.023|3^1D_2\rangle +0.034|3^3D_2\rangle\end{array} $& 6.142 & &            \\
            \cline{2-7}
$(b\bar{s})$& & $3^-$ & $\begin{array}{cc}&0.998|1^3D_3\rangle -0.004|1^3G_3\rangle+0.061|2^3D_3\rangle +0.003|2^3G_3\rangle\\
            &-0.031|3^3D_3\rangle -0.002|3^3G_3\rangle\end{array}$ & 6.144 &&  \\ \cline{2-2}\cline{4-7}
&              &       & $\begin{array}{cc}&0.004|1^3D_3\rangle +0.9995|1^3G_3\rangle-0.005|2^3D_3\rangle -0.024|2^3G_3\rangle\\
            &-0.003|3^3D_3\rangle +0.018|3^3G_3\rangle\end{array}$  & 6.641 & &            \\
            \cline{2-7}
&$ $ & $3^+$ & $\begin{array}{cc}&0.749|1^1F_3\rangle -0.662|1^3F_3\rangle+0.011|2^1F_3\rangle -0.007|2^3F_3\rangle\\
            &-0.009|3^1F_3\rangle +0.007|3^3F_3\rangle\end{array}$ & 6.374 &  &   \\ \cline{2-2}\cline{4-7}
&              &       & $\begin{array}{cc}&0.662|1^1F_3\rangle +0.749|1^3F_3\rangle-0.014|2^1F_3\rangle -0.019|2^3F_3\rangle\\
            &+0.011|3^1F_3\rangle +0.014|3^3F_3\rangle\end{array}$  & 6.416& &            \\
            \cline{2-7}
& & $4^+$ & $\begin{array}{cc}&0.9992|1^3F_4\rangle -0.002|1^3H_4\rangle+0.033|2^3F_4\rangle +0.002|2^3H_4\rangle\\
            &-0.019|3^3F_4\rangle-0.001|3^3H_4\rangle\end{array}$ & 6.380 &  &   \\ \cline{2-2}\cline{4-7}
&              &       & $\begin{array}{cc}&0.002|1^3F_4\rangle +0.9999|1^3H_4\rangle-0.002|2^3F_4\rangle -0.006|2^3H_4\rangle\\
            &-0.001|3^3F_4\rangle +0.010|3^3H_4\rangle\end{array}$  &6.849 & &            \\
                        \cline{2-7}
& & $4^-$ & $\begin{array}{cc}&0.744|1^1G_4\rangle -0.668|1^3G_4\rangle+0.001|2^1G_4\rangle +0.001|2^3G_4\rangle\\
            &-0.005|3^1G_4\rangle +0.004|3^3G_4\rangle\end{array}$ & 6.585 &  &   \\ \cline{2-2}\cline{4-7}
&              &       & $\begin{array}{cc}&0.668|1^1G_4\rangle +0.744|1^3G_4\rangle-0.001|2^1G_4\rangle -0.003|2^3G_4\rangle\\
            &+0.006|3^1G_4\rangle +0.007|3^3G_4\rangle\end{array}$  &6.647 & &            \\
\hline
\end{tabular}
\end{table}
\end{center}

\begin{center}
\begin{table}
\caption{Theoretical spectrum of  $(b\bar{q})$ and $(b\bar{s})$ bound states mainly with
the radial quantum number $n=2$.}
 \label{t4}
\begin{tabular}{|c|c|c|c|c|}\hline
     & $J^P$ & Multiplet & Mass (GeV) & GI (GeV) \\
     \hline
  & $0^-$ & $|2^1S_0\rangle$ & 5.836 & 5.90\\ \cline{2-5}
 &$1^-$ & $\begin{array}{cc} &-0.018|1^3S_1\rangle+0.019|1^3D_1\rangle+0.999|2^3S_1\rangle-0.016|2^3D_1\rangle \\
     &+0.022|3^3S_1\rangle+0.014|3^3D_1\rangle\end{array}$& 5.878 & 5.93  \\ \cline{3-5}
 &      & $\begin{array}{cc} &-0.034|1^3S_1\rangle+0.086|1^3D_1\rangle+0.016|2^3S_1\rangle+0.988|2^3D_1\rangle \\
     &-0.034|3^3S_1\rangle-0.120|3^3D_1\rangle\end{array}$& 6.422 & \\ \cline{2-5}
 & $0^+$ & $|2^3P_0\rangle$ &6.070&  \\ \cline{2-5}
  & $1^+$ &$0.081|1^1P_1\rangle +0.118|1^3P_1\rangle+0.552|2^1P_1\rangle+0.822|2^3P_1\rangle$ & 6.137 &  \\ \cline{3-5}
 &       & $-0.036|1^1P_1\rangle +0.025|1^3P_1\rangle+0.829|2^1P_1\rangle-0.557|2^3P_1\rangle$& 6.196 & \\ \cline{2-5}
 & $2^+$ &  $\begin{array}{cc}& -0.080|1^3P_2\rangle +0.009|1^3F_2\rangle+0.983|2^3P_2\rangle -0.007|2^3F_2\rangle\\
            &+0.168|3^3P_2\rangle +0.006|3^3F_2\rangle\end{array}$ &6.220 & \\ \cline{3-5}
  &       &  $\begin{array}{cc}&-0.007|1^3P_2\rangle +0.022|1^3F_2\rangle+0.012|2^3P_2\rangle +0.998|2^3F_2\rangle\\
            &-0.031|3^3P_2\rangle -0.044|3^3F_2\rangle\end{array}$ & 6.676 &  \\
            \cline{2-5}
 $(b\bar{q})$ & $2^-$ &  $\begin{array}{cc}&-0.011|1^1D_2\rangle +0.004|1^3D_2\rangle+0.753|2^1D_2\rangle -0.657|2^3D_2\rangle\\
            &+0.024|3^1D_2\rangle -0.014|3^3D_2\rangle\end{array}$ &6.417 & \\ \cline{3-5}
  &       &  $\begin{array}{cc}&0.018|1^1D_2\rangle +0.026|1^3D_2\rangle+0.656|2^1D_2\rangle +0.752|2^3D_2\rangle\\
            &-0.034|3^1D_2\rangle -0.048|3^3D_2\rangle\end{array}$ & 6.446 &  \\
            \cline{2-5}
 & $3^-$ &  $\begin{array}{cc}&-0.037|1^3D_3\rangle +0.003|1^3G_3\rangle+0.996|2^3D_3\rangle -0.004|2^3G_3\rangle\\
            &+0.076|3^3D_3\rangle +0.003|3^3G_3\rangle\end{array}$ & 6.430 & \\ \cline{3-5}
  &       &  $\begin{array}{cc}&-0.003|1^3D_3\rangle -0.007|1^3G_3\rangle+0.004|2^3D_3\rangle +0.9999|2^3G_3\rangle\\
            &-0.006|3^3D_3\rangle -0.002|3^3G_3\rangle\end{array}$ & 6.891 &  \\
            \cline{2-5}
 & $3^+$ &  $\begin{array}{cc}&0.006|1^1F_3\rangle -0.007|1^3F_3\rangle+0.751|2^1F_3\rangle -0.660|2^3F_3\rangle\\
            &-0.002|3^1F_3\rangle +0.004|3^3F_3\rangle\end{array}$ & 6.618 & \\ \cline{3-5}
  &       &  $\begin{array}{cc}&-0.007|1^1F_3\rangle -0.006|1^3F_3\rangle+0.660|2^1F_3\rangle +0.751|2^3F_3\rangle\\
            &+0.002|3^1F_3\rangle -0.002|3^3F_3\rangle\end{array}$ & 6.689 &  \\
            \cline{2-5}
 & $4^+$ &  $\begin{array}{cc}&-0.010|1^3F_4\rangle +0.002|1^3H_4\rangle+0.9995|2^3F_4\rangle -0.002|2^3H_4\rangle\\
            &+0.028|3^3F_4\rangle +0.002|3^3H_4\rangle\end{array}$ & 6.626 & \\ \cline{3-5}
  &       &  $\begin{array}{cc}&-0.002|1^3F_4\rangle -0.024|1^3H_4\rangle+0.002|2^3F_4\rangle +0.999|2^3H_4\rangle\\
            &-0.003|3^3F_4\rangle +0.025|3^3H_4\rangle\end{array}$ & 7.085 &  \\

\hline
   & $0^-$ & $|2^1S_0\rangle$ &5.928& 5.98 \\ \cline{2-5}
 &$1^-$ & $\begin{array}{cc} & -0.018|1^3S_1\rangle+0.028|1^3D_1\rangle+0.999|2^3S_1\rangle-0.019|2^3D_1\rangle \\
     &+0.023|3^3S_1\rangle+0.016|3^3D_1\rangle\end{array}$ & 5.970 & 6.01 \\ \cline{3-5}
  &      & $\begin{array}{cc} &-0.014|1^3S_1\rangle+0.132|1^3D_1\rangle+0.018|2^3S_1\rangle+0.976|2^3D_1\rangle \\
     &-0.055|3^3S_1\rangle-0.160|3^3D_1\rangle\end{array}$& 6.460 & \\ \cline{2-5}
 & $0^+$ & $|2^3P_0\rangle$ & 6.109& \\ \cline{2-5}
 & $1^+$ & $0.108|1^1P_1\rangle +0.159|1^3P_1\rangle+0.547|2^1P_1\rangle+0.815|2^3P_1\rangle$ & 6.187 &  \\ \cline{3-5}
 &       & $-0.058|1^1P_1\rangle +0.039|1^3P_1\rangle +0.828|2^1P_1\rangle -0.556|2^3P_1\rangle$ &6.289 & \\ \cline{2-5}
  & $2^+$ &$\begin{array}{cc}& -0.098|1^3P_2\rangle +0.016|1^3F_2\rangle+0.974|2^3P_2\rangle -0.009|2^3F_2\rangle\\
            &+0.202|3^3P_2\rangle +0.007|3^3F_2\rangle\end{array}$ &6.308 &  \\ \cline{3-5}
  &       & $\begin{array}{cc}&-0.024|1^3P_2\rangle +0.056|1^3F_2\rangle+0.055|2^3P_2\rangle +0.965|2^3F_2\rangle\\
            &-0.235|3^3P_2\rangle -0.081|3^3F_2\rangle\end{array}$ & 6.709 & \\
             \cline{2-5}
  $(b\bar{s})$ & $2^-$ &$\begin{array}{cc}&0.046|1^1D_2\rangle +0.057|1^3D_2\rangle+0.561|2^1D_2\rangle +0.818|2^3D_2\rangle\\
            &-0.066|3^1D_2\rangle -0.080|3^3D_2\rangle\end{array}$ &6.483 &  \\ \cline{3-5}
  &       & $\begin{array}{cc}&-0.024|1^1D_2\rangle +0.022|1^3D_2\rangle+0.822|2^1D_2\rangle -0.565|2^3D_2\rangle\\
            &+0.044|3^1D_2\rangle -0.042|3^3D_2\rangle\end{array}$ & 6.499 & \\
              \cline{2-5}
 & $3^-$ &  $\begin{array}{cc}&-0.057|1^3D_3\rangle +0.005|1^3G_3\rangle+0.992|2^3D_3\rangle -0.005|2^3G_3\rangle\\
            &+0.109|3^3D_3\rangle +0.004|3^3G_3\rangle\end{array}$ & 6.512 & \\ \cline{3-5}
  &       &  $\begin{array}{cc}&-0.003|1^3D_3\rangle +0.025|1^3G_3\rangle+0.006|2^3D_3\rangle +0.999|2^3G_3\rangle\\
            &-0.012|3^3D_3\rangle -0.044|3^3G_3\rangle\end{array}$ & 6.919 &  \\
            \cline{2-5}
 & $3^+$ &  $\begin{array}{cc}&-0.011|1^1F_3\rangle +0.007|1^3F_3\rangle+0.740|2^1F_3\rangle -0.672|2^3F_3\rangle\\
            &+0.023|3^1F_3\rangle -0.016|3^3F_3\rangle\end{array}$ & 6.695 & \\ \cline{3-5}
  &       &  $\begin{array}{cc}&0.015|1^1F_3\rangle +0.020|1^3F_3\rangle+0.651|2^1F_3\rangle +0.740|2^3F_3\rangle\\
            &-0.026|3^1F_3\rangle -0.035|3^3F_3\rangle\end{array}$ & 6.722&  \\
            \cline{2-5}
 & $4^+$ &  $\begin{array}{cc}&-0.032|1^3F_4\rangle +0.003|1^3H_4\rangle+0.998|2^3F_4\rangle -0.003|2^3H_4\rangle\\
            &+0.061|3^3F_4\rangle+0.002|3^3H_4\rangle\end{array}$ & 6.704 & \\ \cline{3-5}
  &       &  $\begin{array}{cc}&-0.002|1^3F_4\rangle +0.006|1^3H_4\rangle+0.003|2^3F_4\rangle +0.9998|2^3H_4\rangle\\
            &-0.005|3^3F_4\rangle -0.016|3^3H_4\rangle\end{array}$ & 7.108 &  \\

\hline
\end{tabular}
\end{table}
\end{center}
\end{widetext}


\section*{Acknowledgments} This work is supported in part by the
National Natural Science Foundation of China under contracts Nos.
11375088, 10975077, 10735080, 11125525.



\begin{thebibliography}{99}
\bibitem{Ds2632}A.V. Evdokimov {\it et al.} (SELEX Collaboration), Phys. Rev. Lett. 93, 242001 (2004).
\bibitem{Ds2860}B. Aubert {\it et al.} (BaBar collaboration), Phys. Rev. Lett. 97, 222001 (2006).
\bibitem{Ds2700}J. Brodzicka {\it et al.} (Belle Collaboration), Phys. Rev. Lett. 100, 092001 (2008).
\bibitem{Ds2009}B. Aubert {\it et al.} (BaBar Collaboration), Phys. Rev. D 80, 092003 (2009).
\bibitem{D2010}P. del Amo Sanchez {\it et al.} (BaBar Collaboration), Phys. Rev. D 82, 111101 (2010).
\bibitem{LHCb}R. Aaij {\it et al.} (LHCb collaboration), JHEP 1309, 145 (2013).
\bibitem{RGG}A. De R\'{u}jula, H. Georgi and S.L. Glashow, Phys. Rev. D12, 147 (1975).
\bibitem{GI}S. Godfrey and N. Isgur, Phys. Rev. D32, 189 (1985).
\bibitem{RP1}P. Cea, P. Colangelo, G. Nardulli, G. Paiano, and G. Preparata, Phys. Rev. D26, 1157 (1982).
\bibitem{RP2}P. Cea, G. Nardulli, and G. Paiano, Phys. Rev. D28, 2291 (1983).
\bibitem{RP3}P. Cea, P. Colangelo, L. Cosmai and G. Nardulli, Phys. Lett. B206, 691 (1988).
\bibitem{RP4}P. Colangelo, G. Nardulli, M. Pietroni, Phys. Rev. D43, 3002 (1991).
\bibitem{RP5}P. Colangelo, F. De Fazio, M. Ladisa, G. Nardulli, P. Santorelli, A. Tricarico, Eur. Phys. J. C8, 81 (1999).
\bibitem{ymz}M.Z. Yang, Euro. Phys. J. C72, 1880 (2012).
\bibitem{LY}J.B. Liu, M.Z. Yang, JHEP 1407, 106 (2014).
\bibitem{PDG}J. Beringer {\it et al.} (Particle Data Group), Phys. Rev. D86, 010001 (2012).
\bibitem{PE}M. Di Pierro and E. Eichten, Phys. Rev. D64, 114004 (2001).
\bibitem{Cornell1}E. Eichten, K. Gottfried, T. Kinoshita, K.D. Lane, and T.-M. Yan, Phys. Rev. D17, 3090 (1978)
 [Erratum: Phys. Rev. D 21, 313 (1980)].
\bibitem{Cornell2}E. Eichten, K. Gottfried, T. Kinoshita, K.D. Lane, and T.-M. Yan, Phys. Rev. D21, 203 (1980).
\bibitem{higher-charmonium}T. Barnes, S. Godfrey, and E.S. Swanson, Phys. Rev. D72, 054026 (2005).
\bibitem{HQa1}N. Isgur and M. B. Wise, Phys. Lett. B232, 113 (1989).
\bibitem{HQa2}N. Isgur and M. B. Wise, Phys. Lett. B237, 527 (1990).
\bibitem{HQa3}E. Eichten and B. Hill, Phys. Lett. B234, 511 (1990).
\bibitem{HQa4}E. Eichten and B. Hill, Phys. Lett. B243, 427 (1990).
\bibitem{HQa5}B. Grinstein, Nucl. Phys. B339, 253 (1990).
\bibitem{HQa6}H. Georgi, Phys. Lett. B240, 447 (1990).
\bibitem{HQa7}A. F. Falk, B. Grinstein, and M. E. Luke, Nucl. Phys. B357, 185 (1991).
\bibitem{HQb} For a review see: M. Neubert, Int. J. Mod. Phys. A11, 4173 (1996).
\bibitem{GK}S.Godfrey, R.Kokoski, Phys. Rev. D43,1679 (1991).
\bibitem{CFGN}P. Colangelo, F. De Fazio, F. Giannuzzi, and S. Nicotri, Phys. Rev. D86, 054024 (2012).
\bibitem{BR}E. van Beveren, G. Rupp, Phys. Rev. Lett. 97, 202001 (2006).
\bibitem{CFN} P. Colangelo, F. De Fazio and S. Nicotri, Phys. Lett. B642, 48 (2006).
\bibitem{CFNR}P. Colangelo, F. De Fazio, S. Nicotri and M. Rizzi, Phys. Rev. D77, 014012  (2008).
\bibitem{CTLS}E. F. Close, C. E. Thmas, O. Lakhina, E. S. Swanson, Phys. Lett. B647, 159 (2007).
\bibitem{CF1}P. Colangelo, F. De Fazio, Phys. Rev. D81, 094001 (2010).
\bibitem{Godfrey}S. Godfrey, Phys. Rev. D70, 054017 (2004).
\bibitem{BBP}T. Barnes, N. Black and P. R. Page, Phys. Rev. D68, 054014 (2003).
\bibitem{Belle}K. Abe {\it et al.} (Belle Collaboration), Phys. Rev. D69, 112002 (2004).
\end{thebibliography}
\end{document}